\documentclass[conference]{COSAItran}
\COSAIoverridecommandlockouts
\usepackage{cite}
\usepackage{amsmath,amssymb,amsfonts}
\usepackage{algorithmic}
\usepackage{graphicx}
\graphicspath{{Figures/}}
\usepackage{textcomp}
\usepackage{xcolor}
\usepackage[colorlinks]{hyperref}
\def\BibTeX{{\rm B\kern-.05em{\sc i\kern-.025em b}\kern-.08em
    T\kern-.1667em\lower.7ex\hbox{E}\kern-.125emX}}
\begin{document}

\title{Band strutures of hybrid   graphene quantum dots with magnetic flux\\
}

\author{\COSAIauthorblockN{Bouchaib Lemaalem}
\COSAIauthorblockA{\textit{Laboratory of Theoretical Physics} \\
\textit{Faculty of Sciences}\\
 \textit{Choua\"ib Doukkali University}\\
\textit{El Jadida, Morocco} \\
bouchaib.lemaalem@gmail.com}
\and
\COSAIauthorblockN{Youness Zahidi}
\COSAIauthorblockA{\textit{EMAFI} \\
\textit{Polydisciplinary Faculty}\\
\textit{ Sultan Moulay Selimane University}\\
\textit{Khouribga, Morocco} \\
zahidi.youness@gmail.com }
\and
\COSAIauthorblockN{ Ahmed Jellal}
\COSAIauthorblockA{\textit{Laboratory of Theoretical Physics} \\
\textit{Faculty of Sciences}\\
\textit{ Choua\"ib Doukkali University}\\
\textit{El Jadida, Morocco} \\
a.jellal@ucd.ac.ma}

}

\maketitle

\begin{abstract}
We study the band structures of hybrid graphene quantum dots
subject to a magnetic flux and electrostatic potential. The system is consisting of
a circular single layer graphene 
surrounded by an infinite bilayer graphene. By solving the Dirac equation we obtain the solution of the energy spectrum in two regions. For the valley $K$, it is found that the magnetic flux strongly acts by
decreasing  the gap and shifting energy levels away from zero radius with some oscillations, which are note observed for null flux case. As for the valley $K'$, the energy levels rapidly increase when the radius increases. A number of oscillations appeared that is strongly dependent on the values taken by
the magnetic flux.
\end{abstract}

\begin{COSAIkeywords}
Hybrid graphene,  quantum dots, magnetic flux, electrostatic potential.
\end{COSAIkeywords}

\section{Introduction}
Quantum dots (QDs) in   graphene 
 are very small  particles with unique electronic and optical properties
 \cite{b1,b2,b3,b4,b5,b6}. Since they 
   are  highly tunable, then they can serve as  interesting building blocks for materials that might be used to advance a wide range of applications such as  solar cells, medical imaging and quantum computing.
  Because of  the Klein
tunneling effect  and the absence of the gap in the energy spectrum,
  Dirac fermions cannot be confined by electrostatic potentials
  \cite{jel186}. One solution to overcome such situation is to realize 
 QDs for instance using thin single-layer graphene (SLG) strips \cite{jel1810,jel1811}
or nonuniform magnetic fields \cite{jel1812}. Generally, 
 the electronic and optical
properties of fermions in graphene  depend on  
shapes and edges of QDS \cite{Devrimbook2014}.

On the other hand, A-B bilayer graphene (BLG) consists of two SLG sheets
 where
the A and B atoms in different layers are on top of each other
\cite{b17}, called also 
Bernal stacking. The most important interaction
between the two layers is represented by a direct overlap integral
between A and B atoms on top of each other. BLG presents some particularities for instance 
an external electric field, realized by external gate potentials, can induce a tunable band gap in its energy spectrum contrary to SLG.

Very recently a new generation of circular graphene  QDs has been proposed   based on a hybrid system \cite{Mirzakhani2016}. Indeed, it was  demonstrated that charge carriers can be confined in SLG and BLG islands in a hybrid QD-like structure made of SLG-BLG junctions. As a result, it is found that
the energy
levels exhibit characteristics of interface states in addition to the
emergence  of
anti-crossings and closing of the band gap  in the
presence of a bias potential.

 Motivated by the results reported  in \cite{Mirzakhani2016}, we consider a geometry made 
 of a circular SLG QD  subjected to a magnetic flux and surrounded by an infinite BLG sheet as depicted in Fig. \ref{fig}. We solve Dirac equation in both regions and determine in the first stage the eigenspinors. To derive equations governing the energy levels, we use the zigzag boundary conditions at interface. We numerically analyze our results and show that
 %
 the magnetic flux differently acts on the energy levels for both valleys $K$ and $K'$.


\section{ SLG quantum dots--BLG infinite}

\begin{figure}[htbp]
	\centerline{\includegraphics[width=2.3in]{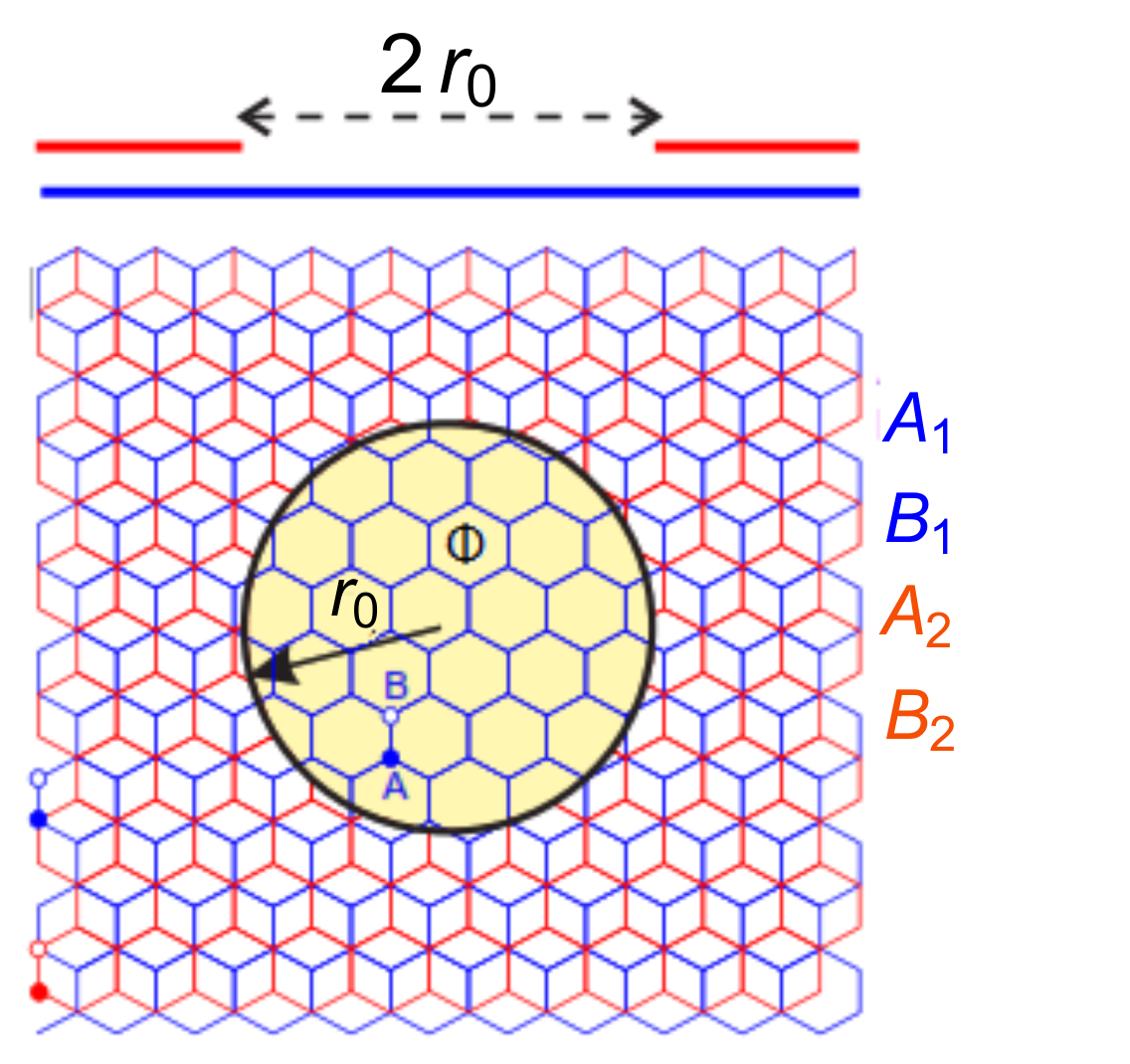}}
	\caption{(color online) Circular SLG quantum dot in the presence of the 
	magnetic flux $\Phi$
 surrounded by an infinite BLGS.}
	\label{fig}
\end{figure}

 We consider a geometry made of 
a circular SLG QD of radius $r_0$ in the presence of a magnetic flux $\phi$
embedded in infinite BLG 
containing $A_1$, $B_1$ in first layer
and $A_2$, $B_2$ in second one (Fig. \ref{fig}). 

The hybrid  system can be described by the following Hamiltonian
\begin{equation}\label{ham1}
H=  v_F\ \vec \sigma\cdot (\vec p + e \vec A) + U_1\mathbb{I}
\end{equation}
where  the vector potential associated to the magnetic flux $\Phi$ is given by
\begin{equation}  A(r)=\frac{h}{e}\frac{\Phi}{2\pi r}\hat{e_{\theta}}
\end{equation}
and $\hat{e_{\theta}}$ is the unit vector for the azimuthal,  $U_1$ is the potential applied to SLG, $\sigma_i$ are Pauli matrices.
To do not couple the two valleys $K$ and $K'$, we assume
 that the spatial extension of the flux
line is large in comparison to the lattice constant.

\subsection{Eigenspinors for the two valleys}

To determine the eigenspinors of fermions in monolayer graphene
subjected to magnetic flux, we consider the basis
$\Psi^{\tau}(r ,\theta)=
(\Psi^{\tau}_A, \Psi^{\tau}_B)^T$ in polar coordinates $(r,\theta)$
because of the system symmetry. The valley index 
$\tau=+1$ refers to   $K$  and $\tau=-1$ to  $K^{'}$.
By showing  that $[H, J_z]=0$, then    the separability imposes 
such as
\begin{equation}
\Psi^{\tau}(r ,\theta)=e^{im\theta}\begin{pmatrix}\chi^{\tau}_A(r) \\ ie^{-i\tau \theta}\chi^{\tau}_B(r)  \end{pmatrix} 
\end{equation}
where $m = 0,\pm1,\pm2,\cdots $ are eigenvalues of  
$L_z$ associated to the total angular momentum $J_z= L_z+\frac{\hbar}{2}\sigma_z$. To proceed further, let us 
 write the Hamiltonian \eqref{ham1} 
\begin{equation}
H= v_F\hbar \begin{pmatrix}
u_1 & 
\pi_+
\\ \pi_- &  u_1
\end{pmatrix}
\end{equation}
 in terms of the operators 
\begin{align}
\pi_{\pm }=e^{\pm i\varphi}\left(-i\frac{\partial}{\partial \xi}\pm\frac{\tau}{\xi}\frac{\partial}{\partial \theta}\pm i\frac{\Phi}{\xi}\right)
\end{align}
such that the change of variable $\xi=\frac{r}{r_0}$ and
the  dimensionless quantities  $\varepsilon=\frac{Er_0}{\hbar v_F}$,  $u_1 =\frac{U_1 r_0}{\hbar v_F}$ are used.

The eigenvalue equation $
H\Psi(\xi ,\theta)=\varepsilon \Psi(\xi ,\theta)$
allows to find two coupled equations
\begin{align}
&
\left(\frac{\partial}{\partial \xi}-\frac{\tau m-1 +\Phi}{\xi} 
\right)\chi^{\tau}_B=\left(\varepsilon-u_1\right)\chi^{\tau}_A\label{R-b}\\
&
\left(\frac{\partial}{\partial \xi}+\frac{\tau m+ \Phi}{\xi}
\right)\chi^{\tau}_A=-\left(\varepsilon-u_1\right)\chi^{\tau}_B \label{R+b}
\end{align}
By injecting \eqref{R+b} into \eqref{R-b} we obtain a second order differential equation for $\chi_A^{\tau}$
\begin{align}
\left[\frac{\partial^2}{\partial \xi^2}+\frac{1}{\xi}\frac{\partial}{\partial \xi}-\frac{1}{\xi^2}\left(\tau m+\Phi\right)^2+ \left(\varepsilon-u_1\right)^2\right]\chi^{\tau}_A=0
\end{align}
having   the Bessel function as solution
\begin{eqnarray}\label{chiA}
\chi^{\tau}_A(\xi)=N^{\tau}J_{\nu}(\lambda\xi)
\end{eqnarray}
where we have set 
 $\lambda=\varepsilon - u_1$
and a new quantum number $\nu =\tau m+\Phi$ depending on
the valley index $\tau$ and flux $\Phi$, $N^{\tau}$ is a constant of normalization. Now replacing \eqref{chiA} in \eqref{R+b}, we  end up with the
second component 
\begin{align}
\chi^{\tau}_B(\xi)=-\tau N^{\tau}J_{\nu-\tau}(\lambda\xi)
\end{align}
With the help of some  relations between Bessel functions, we finally obtain
the eigenspinors in monolayer graphene, which are
\begin{equation}
\Psi^{\tau}(\xi ,\theta)=e^{im\theta}\begin{pmatrix}N^{\tau}J_{\nu}(\lambda\xi) \\ ie^{i\tau \theta}\tau N^{\tau}J_{\nu+\tau}(\lambda\xi)  \end{pmatrix} 
\end{equation}

To achieve our task we consider  the  BLG region described by
Hamiltonian for the valley $K$ $(\tau=+1)$
\begin{equation}
H^+= \begin{pmatrix}
u_++u_- & p_+&t^{'}&0
\\ p_- &  u_0+\delta &0&0\\
t^{'}&0&u_0-\delta& p_-\\
0&0& p_+&u_0-\delta
\end{pmatrix} 
\end{equation}
in the basis $(\Phi_{A_1}^+,\Phi_{B_1}^+, \Phi_{A_2}^+, \Phi_{B_2}^+)$
and 
we have set
$p_\pm= p_x\pm ip_y$, $t^{'}=\frac{tr_0}{\hbar v_F}$, $u_+ = \frac{u_1 + u_2 }{2}$, $u_- =\frac{u_1-u_2}{2}$, and
$u_{1,2} = \frac{U_{1,2} r_0}{\hbar v_F}$, with $U_{1,2}$ 
are the potentials at the two
layers. In this region, the eigenspinors can be decoupled as
\begin{equation}
\Phi^{+}(\xi ,\theta)=e^{im\theta} \begin{pmatrix}\chi^+_{A_1}(\xi) \\ ie^{-i\theta}\chi ^+_{B_1}(\xi) \\ \chi ^+_{B_2}(\xi)\\
ie^{i\theta}\chi ^+_{A_1}(\xi) \end{pmatrix} 
\end{equation}
and then the eigenvalue equation gives rise to the set
\begin{align}
&\left(\frac{d}{d\xi}-\frac{m-1}{\xi}\right)\chi ^+_{B_1}=
\left(\varrho- u_-  \right)\chi ^+_{A_1}-t{'}\chi ^+_{B_2}\\
&\left(\frac{d}{d\xi}+\frac{m}{\xi}\right)\chi ^+_{A_1}=-\left(\varrho- u_- \right)\chi ^+_{B_1}
\\
&\left(\frac{d}{d\xi}+\frac{m+1}{\xi}\right)\chi ^+_{A_2}=\left(\varrho+ u_-  \right)\chi ^+_{B_2}(\rho)-t{'}\chi ^+_{A_1}
\\
&\left(\frac{d}{d\xi}-\frac{m}{\xi}\right)\chi ^+_{B_2}=-\left(\varrho+ u_-  \right)\chi ^+_{A_2}
\end{align}
with $\varrho=\epsilon- u_+$.
These equations can
be decoupled to obtain for instance the following one for $\chi ^k_{A_1}$
\begin{equation}
\left(\frac{d^2}{d\xi^2}+\frac{1}{\xi}\frac{d}{d\xi}-\frac{m^2}{\xi^2}-\mu^2_{\pm }\right)\chi ^+_{A_1}=0
\end{equation}
showing the
eigenvalues 
\begin{equation}\label{ener}
\mu_{\pm }=\left[-\left(\varrho^2+u_-^2\right)\pm\left[\left(\varrho^2-u_-^2\right)t^{'2}+4\varrho^2u_-^2\right]^{\frac{1}{2}}\right]^{\frac{1}{2}}
\end{equation}
and the appropriate solutions vanishing at
$r \longrightarrow \infty$ is
  the modified Bessel function of the
second kind $K_m^\pm=K_m (\mu_{\pm }\xi)$, then we have
\begin{equation}
\chi ^+_{A_1}(\xi)=N^+_1K_m^++N^+_2K_m^-
\end{equation}
This can be used to derive the remaining components of the eigenspinors
as follows 
\begin{align}
&
\chi ^+_{B_1}=
\frac{1}{\varrho-u_-}\left[N_1^+\mu_+K_{m-1}^++N_2^+\mu_-K_{m-1}^-\right]\\
&
\chi ^+_{B_2}=\frac{1}{(\varrho-u_-)t^{'}}
\left[N_1^+\eta_+ K_{m}^+
+N_2^+\eta_-K_{m}^-\right]
\\
&
\chi ^+_{A_2}=\frac{1}{(\varrho^2-u_-^2)t^{'}} \left[N_1^+\mu_+\eta_+K_{m+1}^++
N_2^+\mu_-\eta_-K_{m+1}^-\right]
\end{align}
where we have defined
\begin{equation}
\eta_\pm=(\varrho-u_-)^2+\mu_\pm^2
\end{equation}
$N^+_1$ and $N^+_2$ are two constants of normalization. It is showed
that \eqref{ener} possess a gap in the energy spectrum \cite{1717}
\begin{equation}
\Delta= \frac{\Delta U }{\sqrt{1 +\frac{\Delta U^2}{t^2}}}
\end{equation}
resulted from the Mexican-hat  shaped low-energy dispersion in pristine BLG. In the limit $\Delta U\ll t$ it behaves as potential scale,
i.e. $\Delta\approx {\Delta U }$.

As concerning the valley $K'$, 
one can easy show that the radial parts of
the corresponding eigenspinors can be linked to
those for the valley $K$. Indeed, solving
the eigenvalue equation
with the basis
\begin{equation}
\Phi^{-}(\xi ,\theta)=e^{im\xi} \begin{pmatrix}\chi ^{-}_{A_1}(\xi)\\ ie^{i\theta}\chi ^{-}_{B_1}(\xi) \\
\chi ^{K^{'}}_{B_2}(\xi)\\
ie^{-i\theta}
\chi ^{-}_{A_2}(\xi) \end{pmatrix} 
\end{equation}
to end up with the relations
\begin{align}
&\chi ^{-}_{A_1}(\xi)=\chi ^{+}_{B_2} (\xi)\\
&\chi ^{-}_{B_1}(\xi) = \chi ^{+}_{A_2}(\xi)\\
&\chi ^{-}_{B_2}(\xi)=\chi ^{+}_{A_1}(\xi)\\
&\chi ^{-}_{B_1}(\xi)= \chi ^{+}_{B_1}(\xi)
\end{align}
In the next we will see how the above results will be
used to analyze the
 influence of the applied magnetic flux on the energy levels.

 \subsection{Zigzag boundary conditions}
 To derive equations describing the energy levels for the two valleys, 
 we use
 the zigzag boundary conditions \cite{b26}. Then,  at the interface
 $\xi=1$, namely $r=r_0$, between SLG and BLG we have the continuities
\begin{align}
\Psi^{\tau}_A(1,\theta)=\Phi_{A_1}^\tau(1,\theta)\label{bcon1}\\
\Psi^{\tau}_B(1,\theta)=\Phi_{B_1}^\tau(1,\theta)\label{bcon2}\\
0=\Phi_{A_2}^\tau(1,\theta)\label{bcon3}
\end{align}
Consequently,  
  (\ref{bcon1}-\ref{bcon3})  give 
\begin{align}
M^+\begin{pmatrix}N^+ \\ N^+_1 \\ N^+_2 \end{pmatrix} =0
\end{align}
for the valley $K$ ($\tau=+1$)
such that the matrix is
\begin{equation}
M^+=
\begin{pmatrix}-J_\nu (\lambda) &K_m^+&K_m^-\\ J_{\nu-1}(\lambda) &a_+K_{m-1}^+&a_-K_{m-1}^-\\0&b_+K_m^+&b_-K_m^- \end{pmatrix}
\end{equation}
where we have set 
\begin{align}
a_{\pm} = \frac{\mu_{\pm }}{\varrho -u_-},\qquad b_{\pm}=\frac{\left(\varrho -u_-\right)^2+\mu_{\pm }^2}{\left(\varrho -u_-\right)t{'}}
\end{align}
Regarding $K^{'}$, we fix $\tau=-1$ and use the corresponding eigenspinors  to end up with 
\begin{align}
M^{-}\begin{pmatrix}N^- \\ N^-_1 \\ N^-_2 \end{pmatrix} =0
\end{align}
and the matrix reads as
\begin{align}
M^{-}=
\begin{pmatrix}-J_\nu (\lambda) &b_+K_m^+&b_-K_m^-\\ -J_{\nu+1}(a) &c_+K_{m-1}^+&c_-K_{m-1}^-\\0&K_m^+&K_m^- \end{pmatrix} 
\end{align}
with  the parameters 
\begin{equation}
c_{\pm} =\frac{b_{\pm} \mu_{\pm } }{\varrho+u_-}
\end{equation}
$N^-_1$ and $N^-_2$ are two constants of normalization.
At this level we point out that the energy levels for both valleys
are solutions of the two determinants
\begin{align}
\det M^+=0, \qquad \det M^-= 0
\end{align}
It is clear that form the complexity of the special functions
involved in
the eigenspinors, it is not easy to analytically derive
an explicit  expression
of the energy levels. Then, to investigate the basic features
of our hybrid system  
 we will  proceed numerically.

\section{Numerical  Analysis}

 We plot the energy levels $E$(eV) versus the dot radius $r_0$(nm) for the valleys 
 $K$ and  $K^{'}$ under suitable conditions of the physical parameters.
 It is convenient for our task to  choose the angular momenta
 $m = 0,\pm1,\pm2,\pm3 $, magnetic flux $\Phi=\pm \frac{1}{2}, \pm \frac{3}{2}, \pm \frac{5}{2}$ and 
the biased potential
 $U_1=-U_2=0.1$ eV.   We emphasis that the solid black
horizontal lines in all figures show the band gap.

\subsection{Valley $K$}

As a first result regarding  
the valley $K$, we observe that
 the symmetry $E^+(m)=E^+(-m)$ is always  preserved for all values taken by
 the magnetic flux.


\begin{figure}[htbp]
	\centerline{
	\includegraphics[scale=0.225]{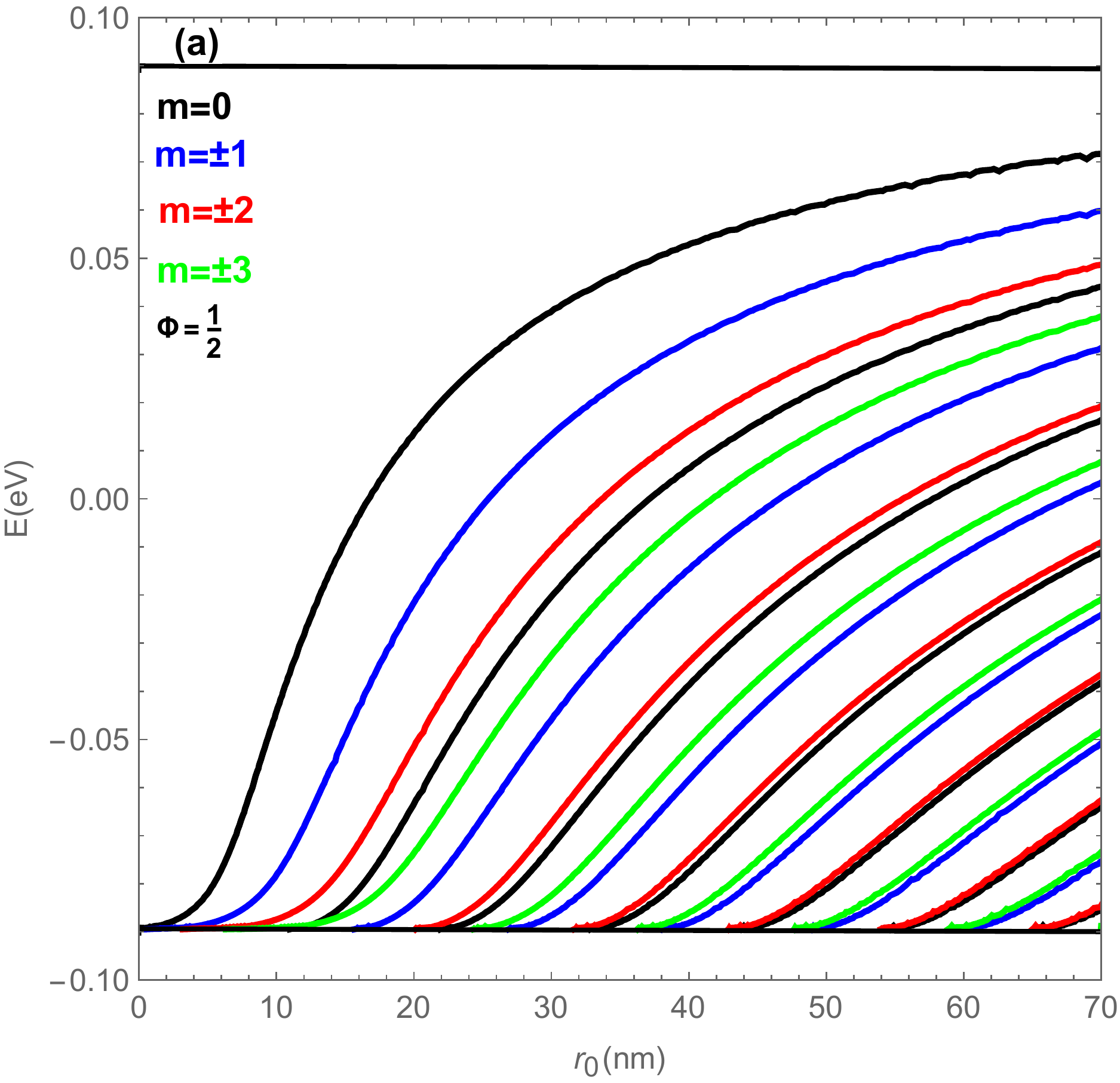}
	\includegraphics[scale=0.225]{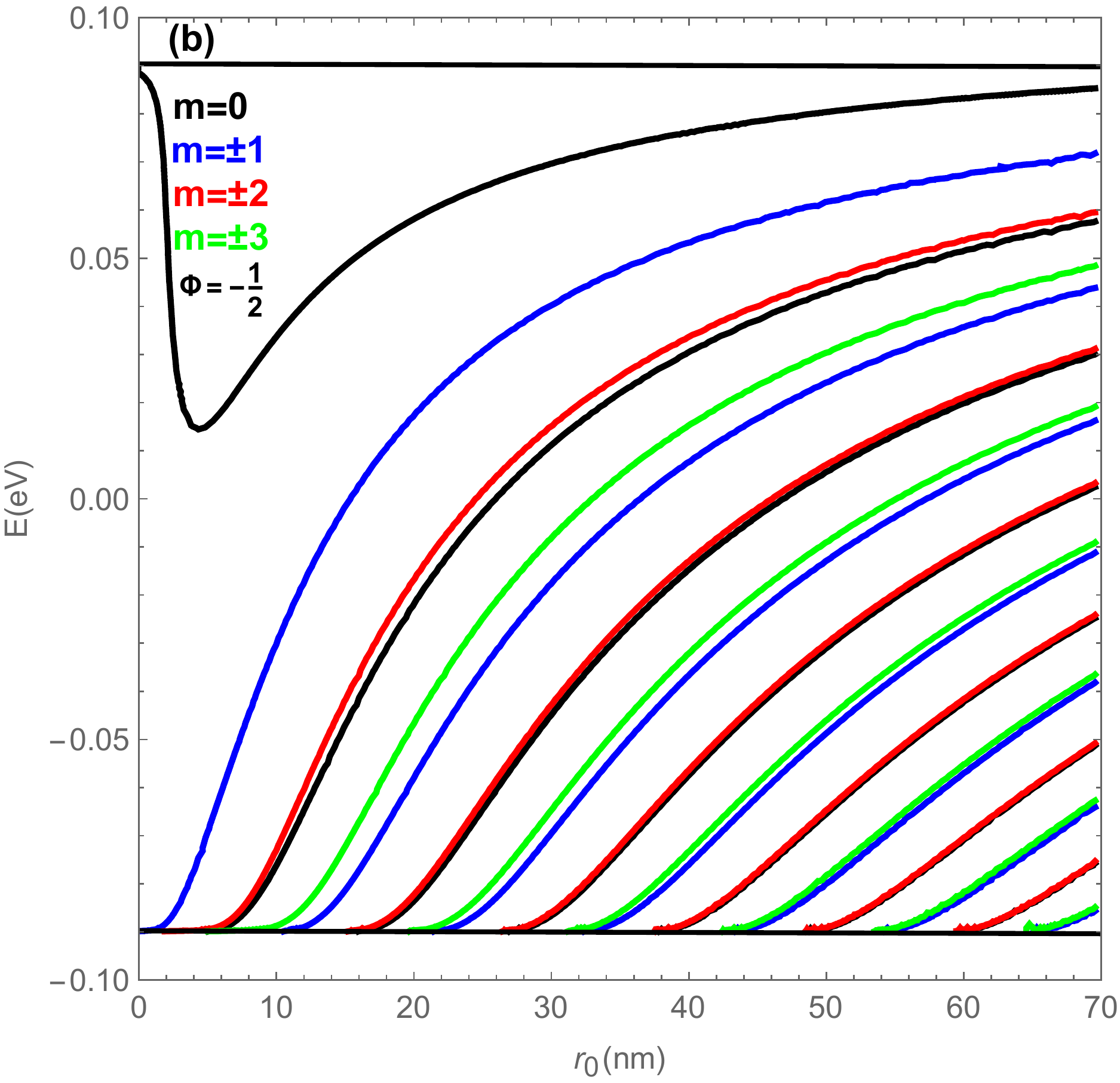}}
	\caption{(color online) Energy levels of the valley $K$
	as a function of the dot radius $r_0$
	 for $m =
		0,\pm1,\pm2,\pm3$  in the presence
		of the biased potential $U_1=-U_2=0.1$ eV and the magnetic flux $\Phi$ (a):
		$\frac{1}{2}$, (b): $-\frac{1}{2}$.
 }\label{Fig12}
\end{figure}


 Fig. \ref{Fig12}a presents the behavior of the energy levels $E$ for the value
 $\Phi= \frac{1}{2}$. We notice that 
 the energy levels  are shifted and band gap decreased compared to the 
 non-flux case \cite{Mirzakhani2016}. As long as the radius increases we see that  
 $E$ increase but gap continue decreasing. We observe another behavior
 in   Fig. \ref{Fig12}b with
 $\Phi= -\frac{1}{2}$. Indeed,
  the energy levels show a maximum corresponds to $m=0$ and $r_0=0$, after  they oscillate
 to reach a minimum at $r_0=4.8$ nm and then increase.


\begin{figure}[htbp]
	\centerline{
	\includegraphics[scale=0.225]{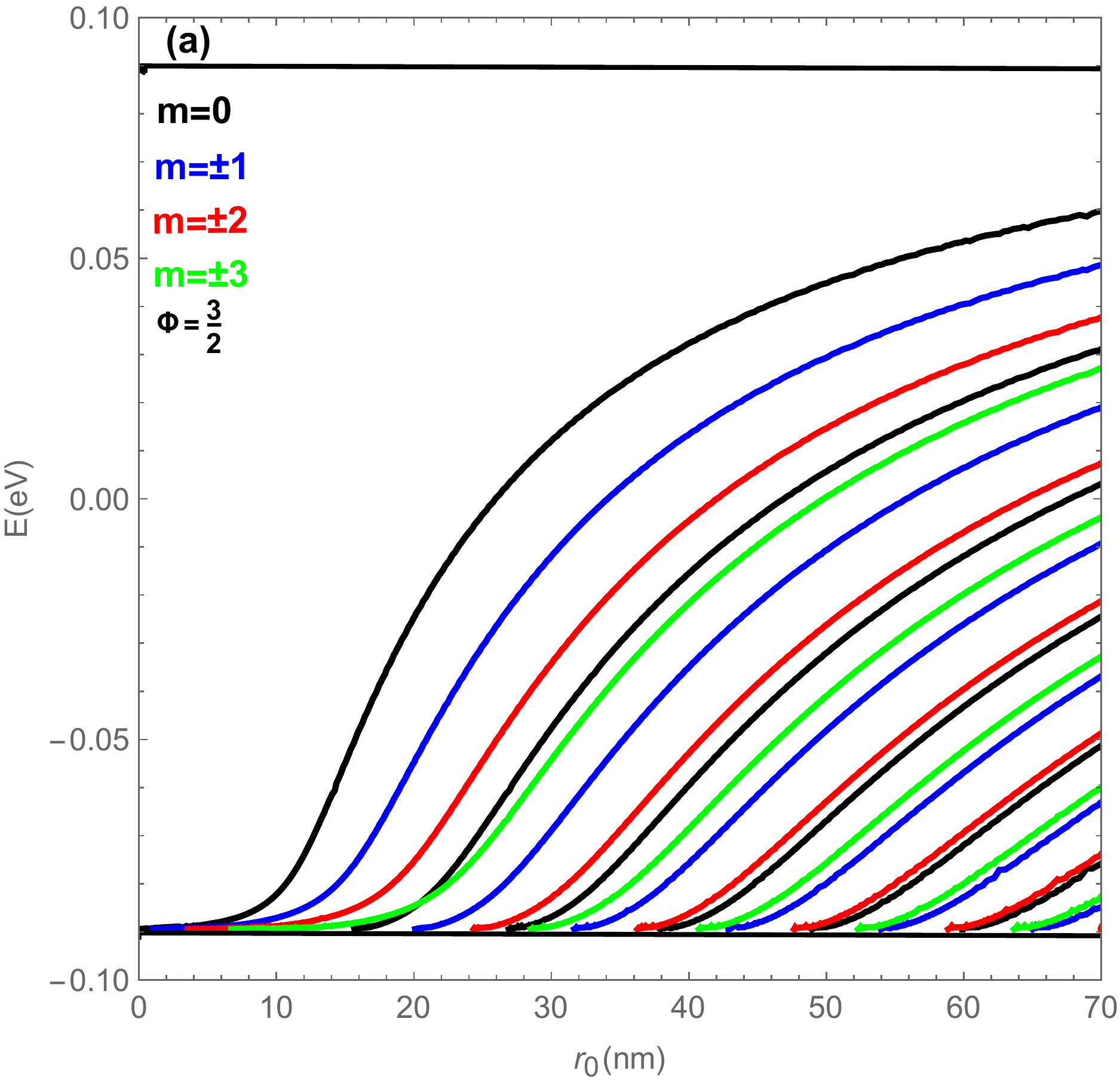}
	\includegraphics[scale=0.225]{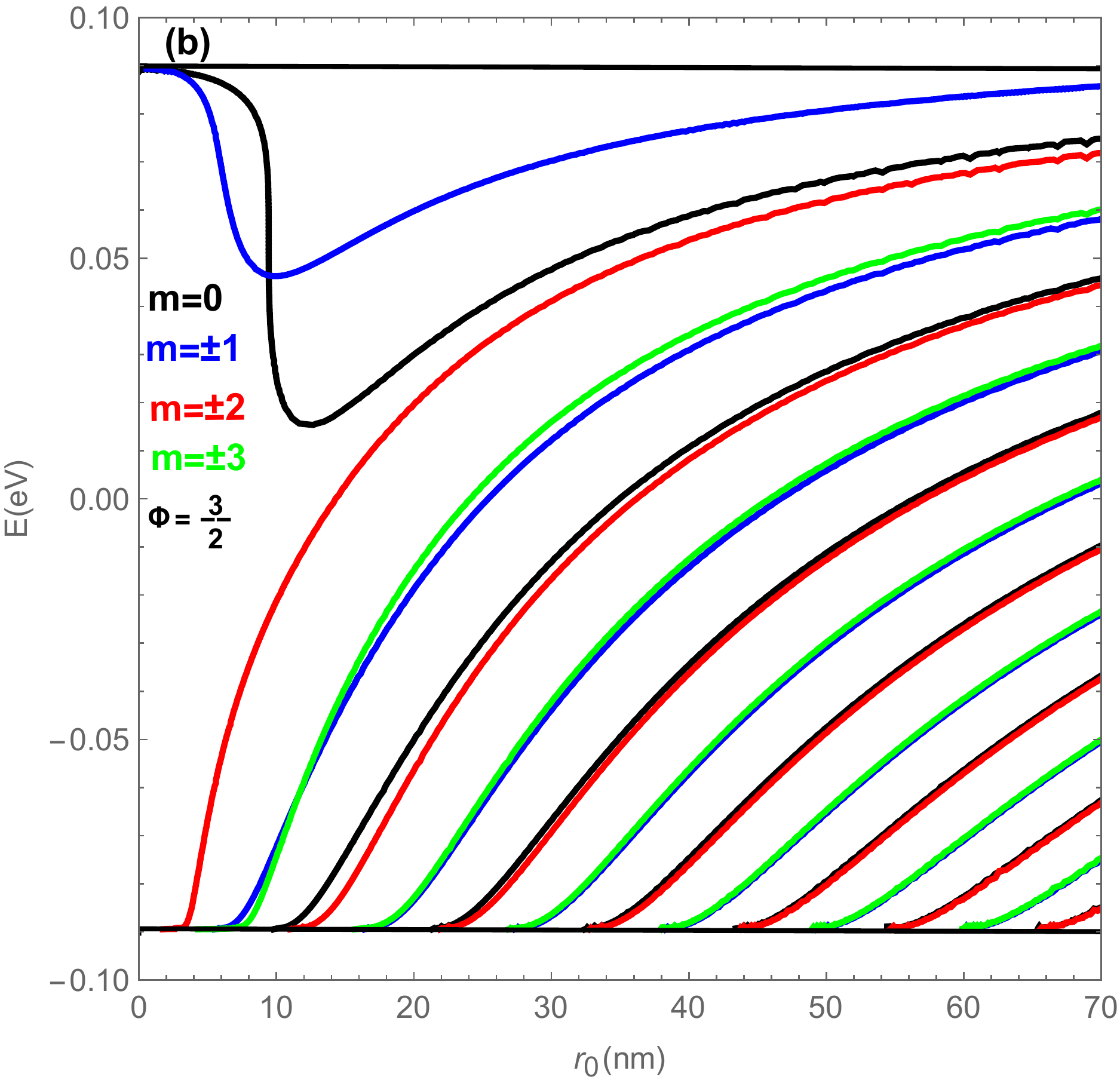}}
	\caption{(color online) The same as in Fig. \ref{Fig12}, but now with the magnetic flux $\Phi$ (a):
		$\frac{3}{2}$, (b): $-\frac{3}{2}$.
 }\label{Fig13}
\end{figure}

Now we increase the magnetic flux and choose  $\Phi= \frac{3}{2}$ as presented in  Fig. \ref{Fig13}b. In the present case, the shift of the energy levels  becomes clear and the decrease in gap as well. Remarkably
 for the negative value $\Phi= -\frac{3}{2}$ in Fig. \ref{Fig13}b we notice that  the values $m=\pm 1$ show
 a similar behavior as in Fig. \ref{Fig12}b for $m=0$ but with less oscillation. In addition, we observe  here that $m=0$ and $m=\pm 1$  correspond to the minima $r_0=12.67$ nm and $r_0=10.11$ nm, respectively.  

\begin{figure}[htbp]
	\centerline{
	\includegraphics[scale=0.225]{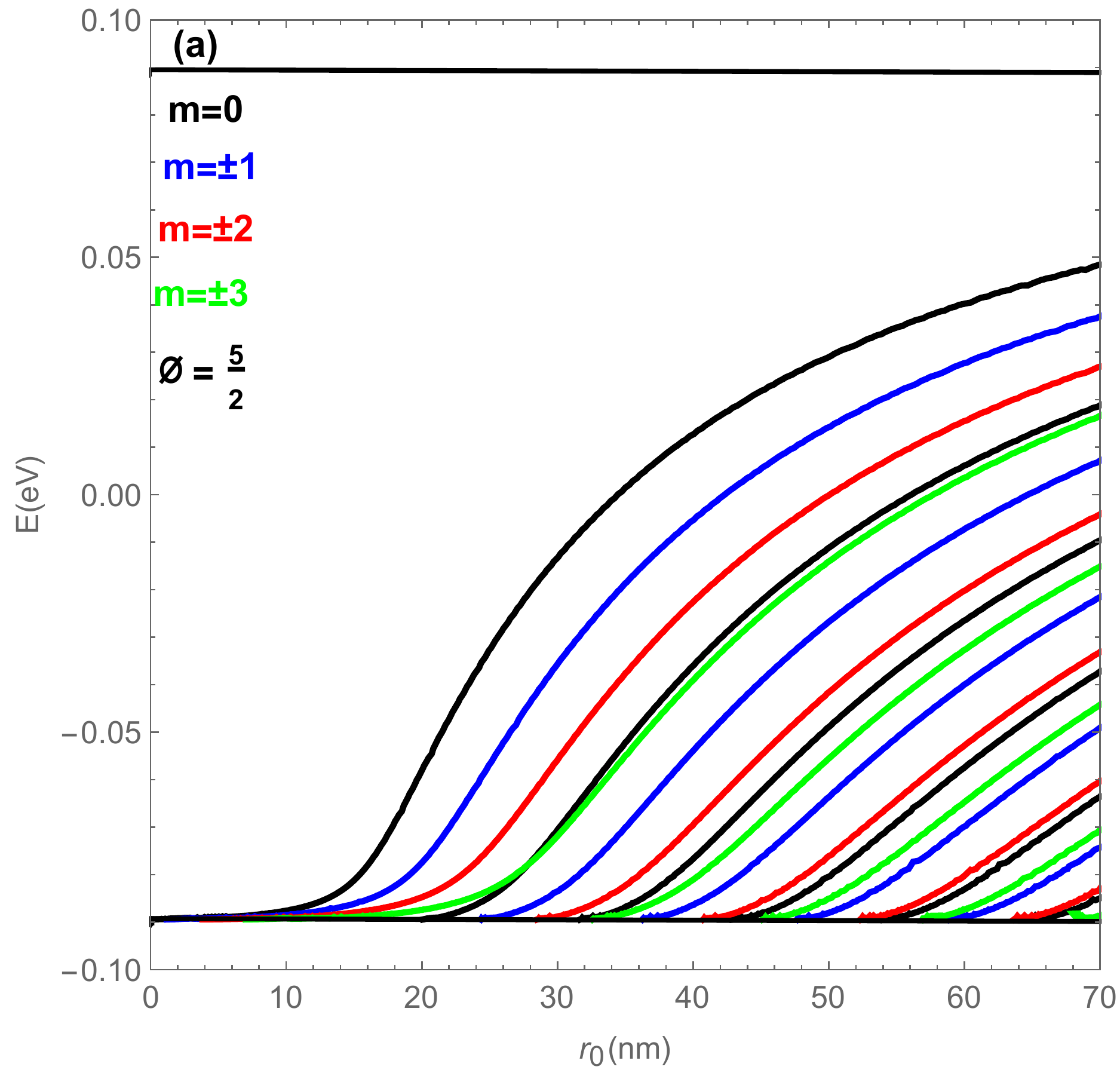}
	\includegraphics[scale=0.225]{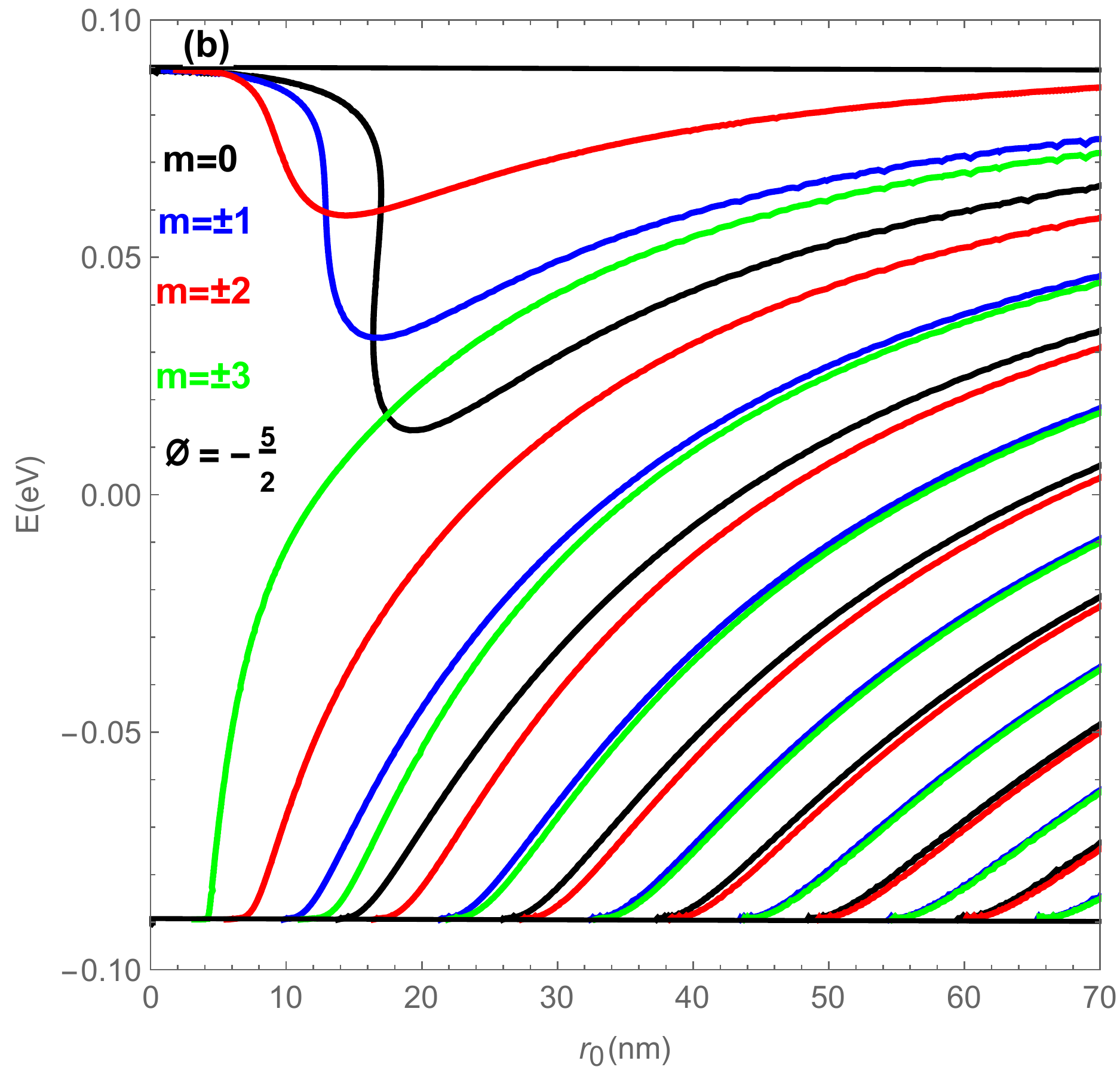}}
	\caption{(color online)  The same as in Fig. \ref{Fig12}, but now the magnetic flux $\Phi$ (a):
		$\frac{5}{2}$, (b): $-\frac{5}{2}$. }\label{Fig15}
\end{figure}

By increasing the magnetic flux to $\phi=\frac{5}{2}$ in
Fig. \ref{Fig15}a, we notice that the shift and gap become more important. As for $\phi=-\frac{5}{2}$ in
Fig. \ref{Fig15}b,  we observe that the number of oscillations increases to three corresponding to the quantum numbers  $m=0$,  $m=\pm 1$ and $m=\pm 2$ with the minima 
$r_0=18.81$ nm,  $r_0=15.69$ nm and $r_0=13.67$ nm, respectively.

\subsection{Valley $K'$}

\begin{figure}[htbp]
	\centerline{
	\includegraphics[scale=0.225]{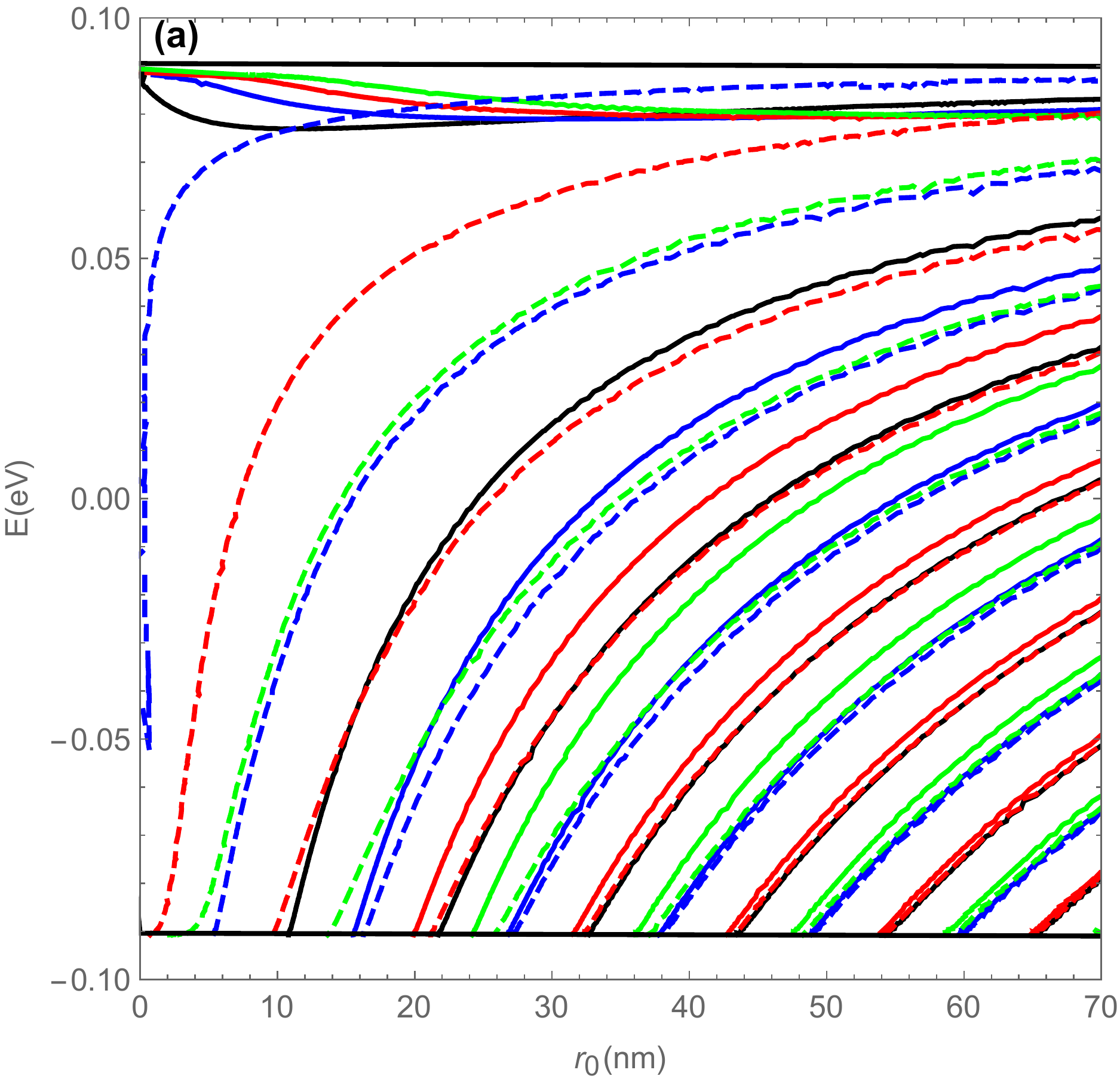}\ \ \  
	\includegraphics[scale=0.225]{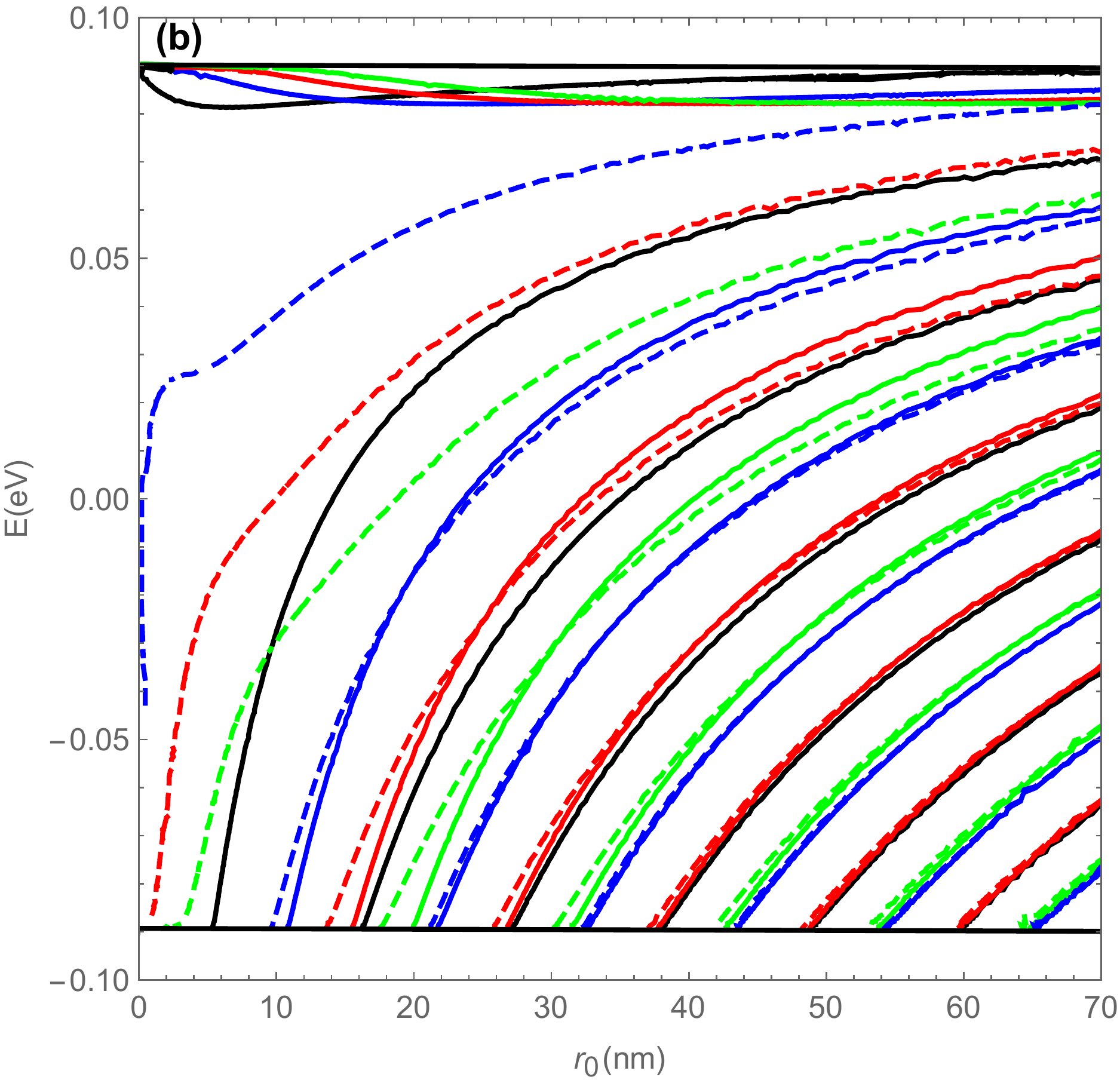}}
	\caption{(color online) Energy levels of the valley $K'$
	as a function of the dot radius $r_0$
	 for $m =
		0,\pm1,\pm2,\pm3$  in the presence
		of the biased potential $U_1=-U_2=0.1$ eV and the magnetic flux $\Phi$ (a):
		$\frac{1}{2}$, (b): $-\frac{1}{2}$.
 }\label{Fig-12}
\end{figure}

As for the valley $K'$, the energy levels show different behavior
resulted form the choice of the magnetic flux compared to the valley $K$. Indeed, 
the first difference that should be noticed we do not have
the symmetry 
$E^-(m)\neq E^-(-m)$ with $m=0,1,2,3$ correspond to solid lines
and the negative numbers to dashed lines in  plots below. In addition,
for $\Phi= \frac{1}{2}$ in Fig. \ref{Fig-12}a, the energy levels rapidly increase and the particularity is seen for $m=-1$ where 
the level overcomes that for $m=1$. The situation is completely changed
for $\Phi= -\frac{1}{2}$ in Fig. \ref{Fig-12}b because we observe
that the level $m=-1$ is under $m=1$.

\begin{figure}[htbp]
	\centerline{
	\includegraphics[scale=0.225]{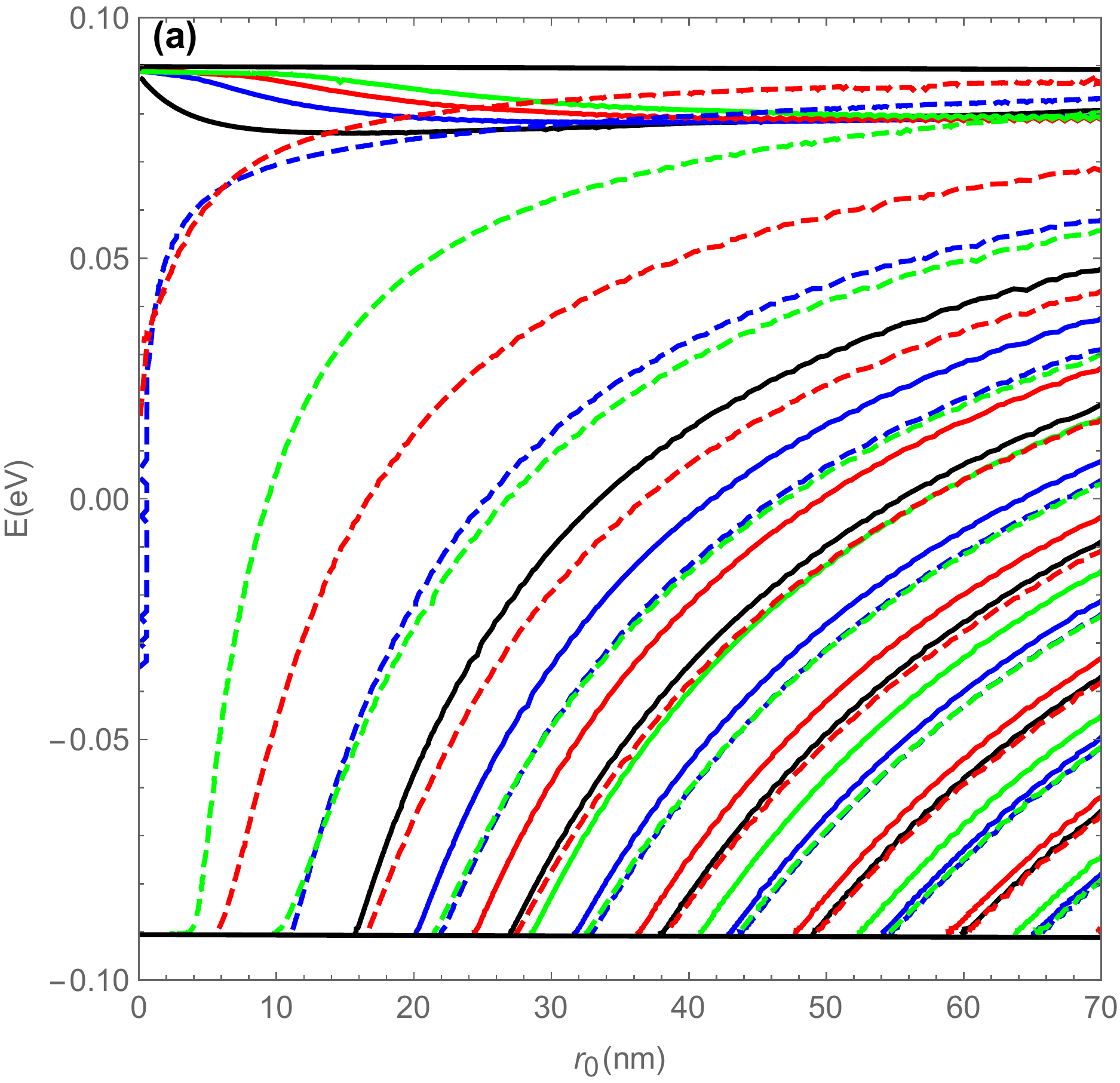}
	\includegraphics[scale=0.225]{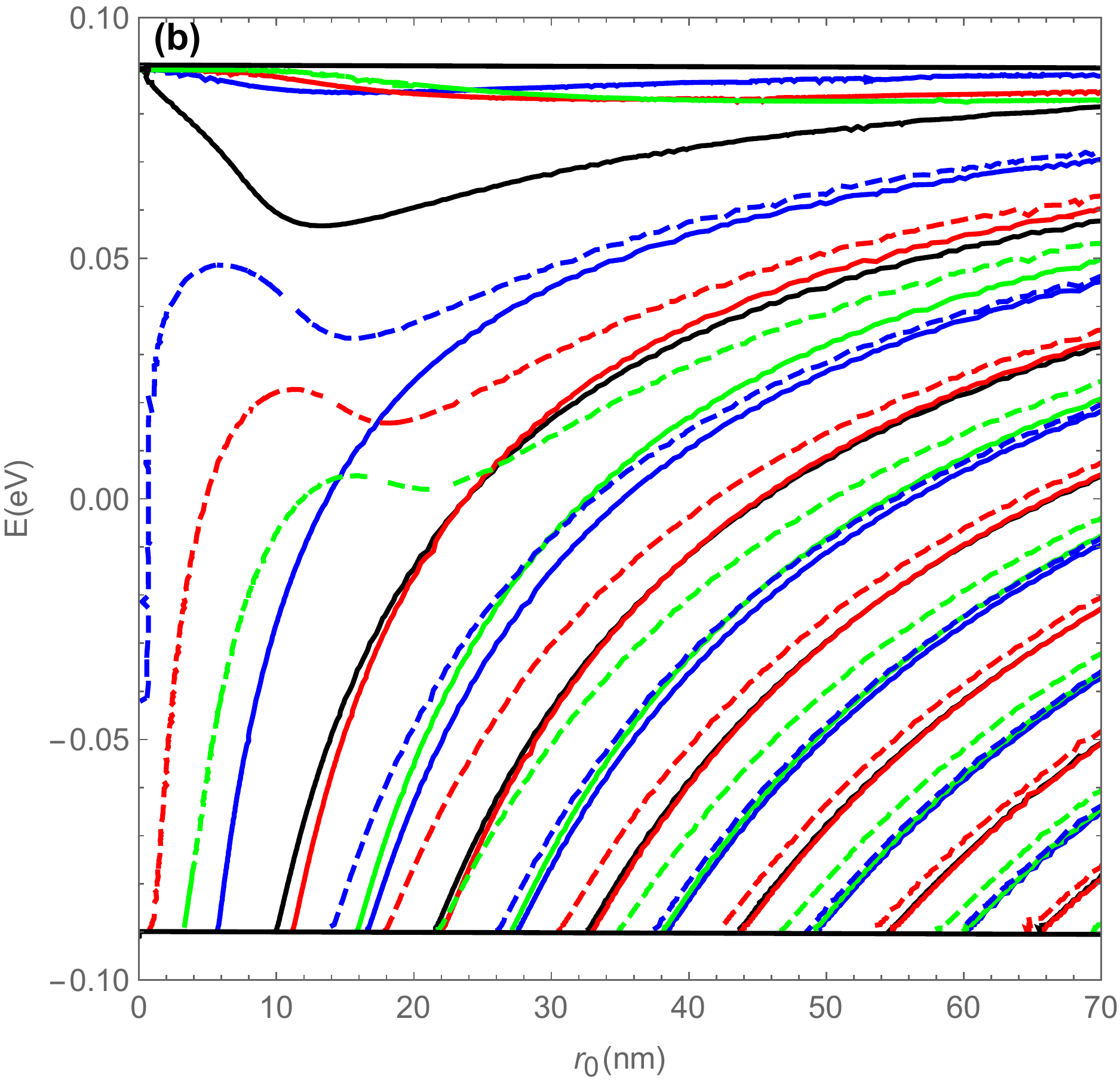}}
	\caption{(color online) The same as in Fig. \ref{Fig-12}, but now with the magnetic flux $\Phi$ (a):
		$\frac{3}{2}$, (b): $-\frac{3}{2}$.
 }\label{Fig-13}
\end{figure}

The behavior of the energy levels becomes more clear in Fig. \ref{Fig-13}a
with $\Phi=\frac{3}{2}$ compared to  Fig. \ref{Fig-12}a. Now as concerning the negative value
  $\Phi= -\frac{3}{2}$, 
Fig. \ref{Fig-13}b shows a mixing of the energy levels
and we observe that some oscillations with different amplitudes start to take place. These are corresponding to
the quantum numbers $m=0$ (black solid line), $m=-1$ (blue dashed line), $m=-2$ (red dashed line), $m=-3$ (green dashed line). 

\begin{figure}[htbp]
	\centerline{
	\includegraphics[scale=0.225]{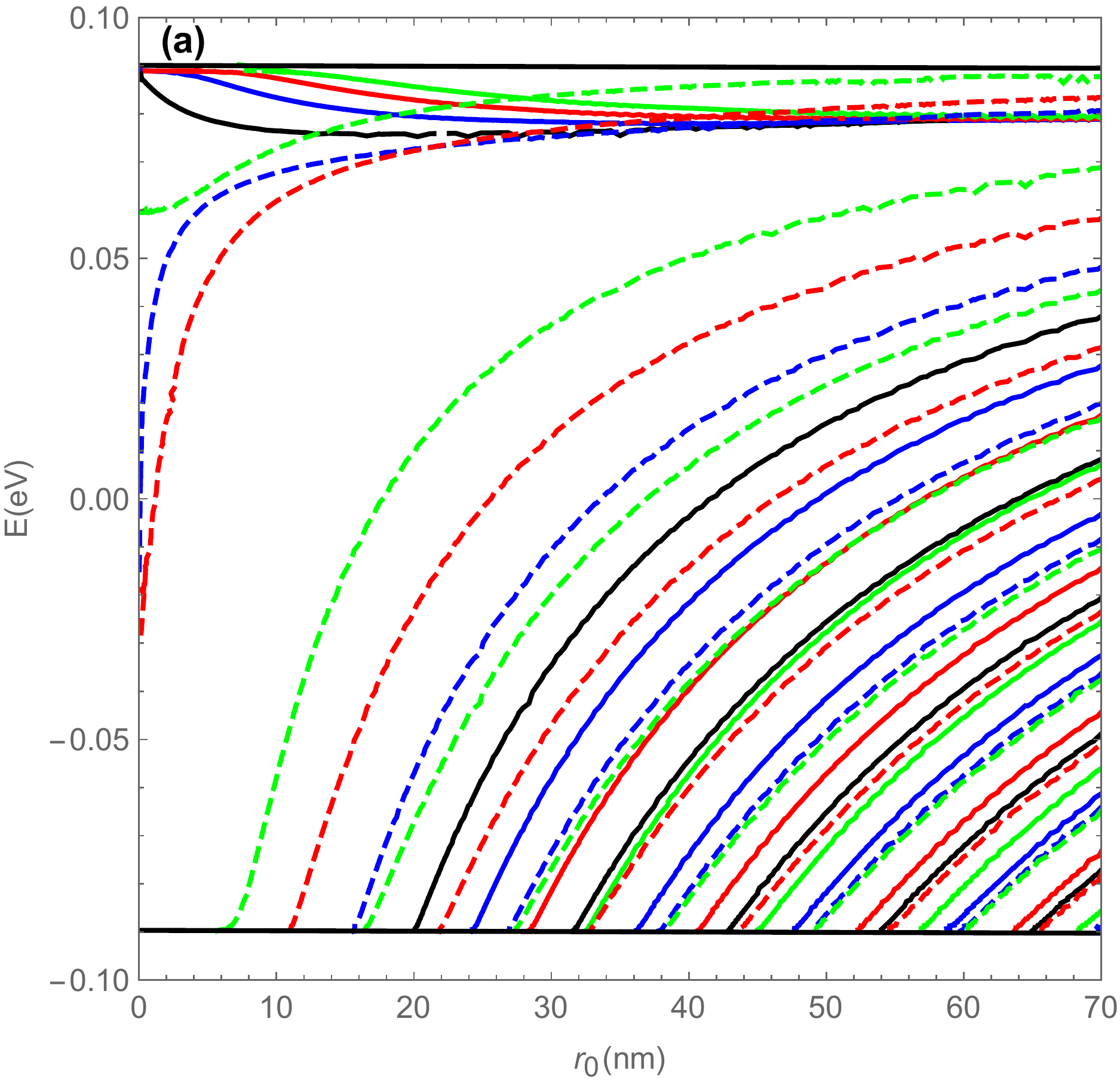}
	\includegraphics[scale=0.225]{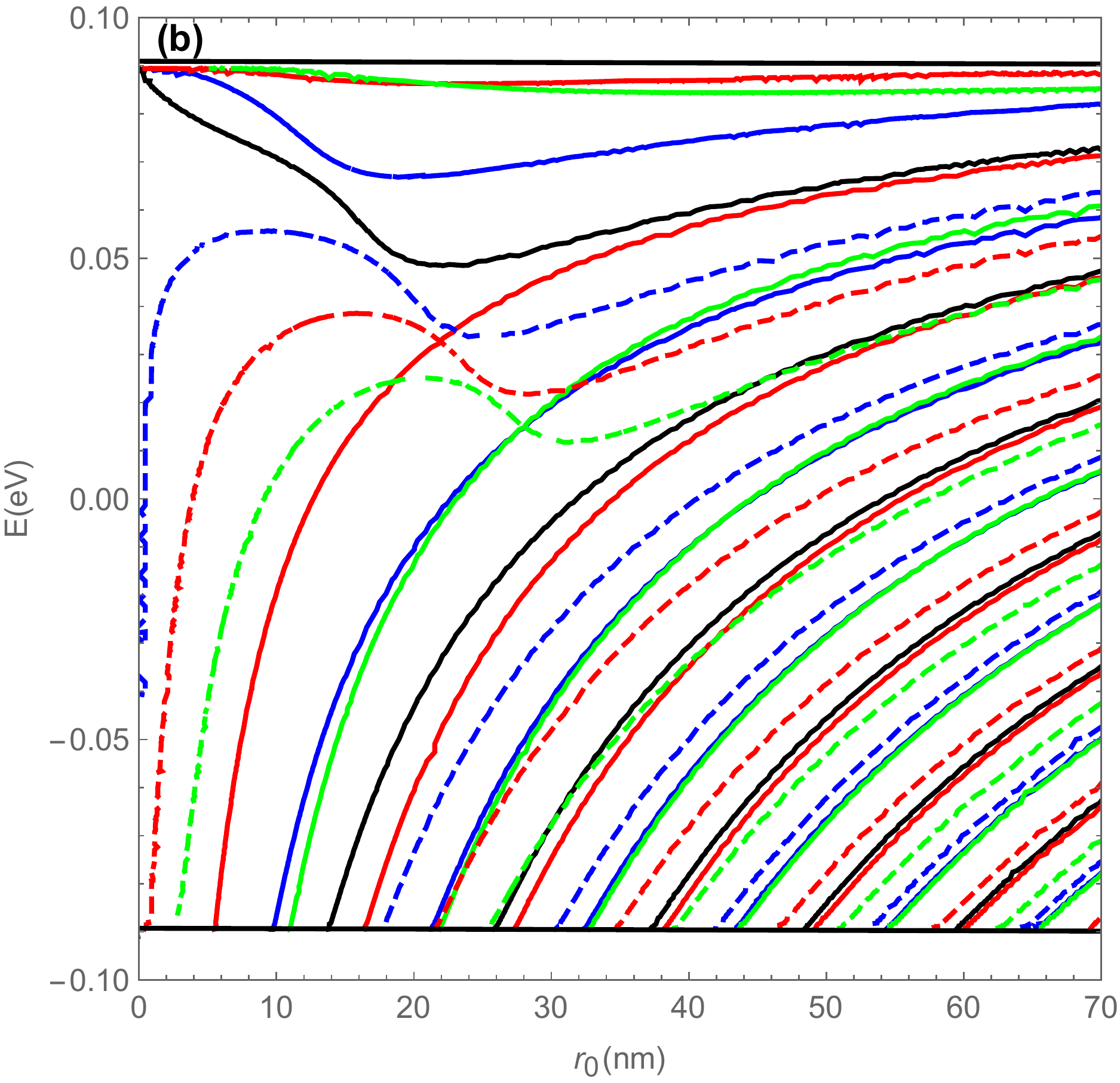}}
	\caption{ (color online) The same as in Fig. \ref{Fig-12}, but now with the magnetic flux $\Phi$ (a):
		$\frac{5}{2}$, (b): $-\frac{5}{2}$. }\label{Fig-15}
\end{figure}

We now consider $\Phi=\frac{5}{2}$ in Fig. \ref{Fig-15}a
and observe an important change in the energy levels such that
shifts increase. In Fig. \ref{Fig-15}b with $\Phi=-\frac{5}{2}$,
we notice that the number of oscillations increase as well. These
results show the manifestation of the magnetic flux and its impact on
the energy levels.

\section{Conclusion}

We have studied the influence of a magnetic flux on the energy levels
of a hybrid graphene. More precisely, we have considered a graphene quantum dot surrounded by a infinite bilayer graphene. Our result showed that the energy levels can oscillate, increase and shift under appropriate choice of flux and variation of the dot radius. We have found that the energy levels of the  valleys $K$ and $K'$ present different behaviors.

	\end{document}